\documentclass[pre,aps,preprint,amssymb]{revtex4}

\pdfoutput=1

\usepackage{amsmath}
\usepackage{tabularx}
\usepackage{pifont}
\usepackage{makeidx}
\usepackage{epsfig}
\usepackage{dcolumn}

\begin{document}

\title{The second law of thermodynamics is riddled with (in percentage terms
extremely rare) exceptions}

\date{\today}
\author{Hans R. Moser \\
{\small Physik-Institut, University of Z\"urich,}
{\small Winterthurerstrasse 190, CH-8057 Z\"urich, Switzerland} \\
{\small E-mail: moser@physik.uzh.ch}}

\begin{abstract}
Statistical mechanics descriptions of the second law of thermodynamics
generally imply point-like particles driven by a dissipative overall
mechanism for their simultaneous time-evolution. As the number of
involved particles grows larger, it becomes more and more unlikely that
they by itself adopt an off-equilibrium state with lower entropy.
We present a macroscopic counterexample of the second law that
{\it repeatedly and spontaneously} produces an entropy sink, thus
recurrently enables us to harvest energy that sidesteps all the
compensation interactions with the surroundings. Hence, this mechanism
extracts energy from a single reservoir. This proves true in an experiment
and is explained as a consequence of size effects, among them nonzero
particle extent that marginally amend crucial peculiarities of
thermodynamic equilibrium dynamics.

\vspace{5mm}

\vspace{5mm}

\end{abstract}

\maketitle

\section{Introduction}
Most of the insights in an entropy context have been gained comparatively late
in science history, at least in the version of today's textbooks. In part, this
is because the second law in some respects is counterintuitive if we disregard
the internal microstructure (atoms, molecules) of matter. Only the statistical
nature of thermal energy gives rise to another quantity, namely entropy
that is not a conserved one in the spirit of Noether's theorem but, e.g.,
quantifies the number of possibilities the microscopic constituents of
gases and liquids may be arranged. Interestingly, also entropy contains
"certain aspects" of a conserved quantity: generalized entropies
such as the Kolmogorov entropy are dynamical invariants in systems
$\dot{\bf x}(t) = {\bf F}({\bf x}(t))$ with irregular time-evolution ${\bf x}(t)$.

We start our actual discussion with Ref.~\cite{holian} that provides a helpful
view of Loschmidt's paradox. These authors infer that non-equilibrium situations
which evolve the same way as time reversal would exhibit, although physically
possible, can be observed with zero probability. We fully agree but
supplement a major issue. Second law violations in a sense resemble
embedding of the rational numbers in the reals. The probability to "accidentally
meet" a rational one is precisely zero. Nevertheless, since we know how to
set up rational numbers, it is well possible to list examples. Below
we present an intentional construction of such a rare situation, namely
a second law violation among the infinitely more frequent cases where the
second law holds true.

Often the potential occurrence of second law violations is mapped to
the question whether a so-called Maxwell's demon can exist. Such an agent is
thought to inspect the constituents of, say, a liquid on a microscopic level and
to separate the fast particles from the slow ones. This type of considerations has
a long, doleful history, sometimes due to an unclear situation whether or not
the demon is part of the system under investigation. In our approach we
pursue quite another course, namely we organize a situation where a spontaneous
departure from thermodynamic equilibrium takes place. We achieve
a quasi-equilibrium between Brownian particles and those of their host
medium, and so this marginal deviation from equilibrium is attended with
thermal processes of minor efficiency. Qualitatively, however, we face
a second law exception with all its imperative consequences such as energy
extraction out of a single reservoir.

\section{Experimental setup}
Now we outline the essentials of our experimentally realized device. One of its
major features is a porous medium that acts as a moderate obstacle to Brownian
motion in a liquid. Some of the Brownian particles by chance enter the region
of the obstacle. Then, motion and accessible positions of the particles
within the porous vicinity are subject to geometrical restrictions, leading to
constrained or "frustrated" Brownian motion. We anticipate that this
constitutes an entropy sink where the {\it overall entropy}, i.e., not just
the one of the captured particles with respect to their prior state, gets
reduced if the obstacle (or trap) is suitably tailored. This arrangement entails
a slight departure from thermodynamic equilibrium of all the liquid (particles of
the host medium plus Brownian ones). The larger Brownian particles transfer
a small fraction of their kinetic equilibrium energy to their immediate
neighborhood that this way marginally warms up. We harvest this "extra energy"
beyond equilibrium, and we postpone a description how to organize this in
a recurrent manner. The physical background on Brownian fluctuations as well as
thermodynamics in general may be found in textbooks \cite{kubo,blundell}.
For entropic consequences if particles enter a porous medium, see~\cite{reguera}.
Further, Brownian particles in confined cavities are an ongoing research subject,
Ref.~\cite{wang} presents a recent approach. But, to our knowledge, related
possible limits of the second law's validity have never been tackled, or at most
in a rather indirect manner.

Figure~1 provides a schematic view of the setup that, at first glance, appears
rather simple. A plug of glass fibers, as they commonly get used
for filtering purposes in chemistry, is sandwiched between two so-called
thermoelectric generators (TEGs). On the two narrow sides of this
device (front, rear in figure), we place glass plates (not shown) that
limit the plug laterally. The TEGs are squares of 30~mm side length
and contain 127 semiconductor thermocouples connected in series, each.
Thus, the Seebeck effect produces a voltage if the two faces of
a TEG are at different temperatures. The thickness of the plug
of glass wool (spacing of the TEGs) amounts to 7~mm and the TEGs
are some 6~mm immersed into saturated solution of potassium permanganate
in water at room temperature. The clearance of the TEGs down to the bottom
of the shallow container with the liquid is around 2~mm (whereas the plug
stands right on the bottom). These geometrical settings prove quite
interdependent in their impact on our measured outcome, thus any change involves
a new overall optimization.

\begin{figure}

\vspace{-1.8cm}

\includegraphics[scale=0.68, angle=0]{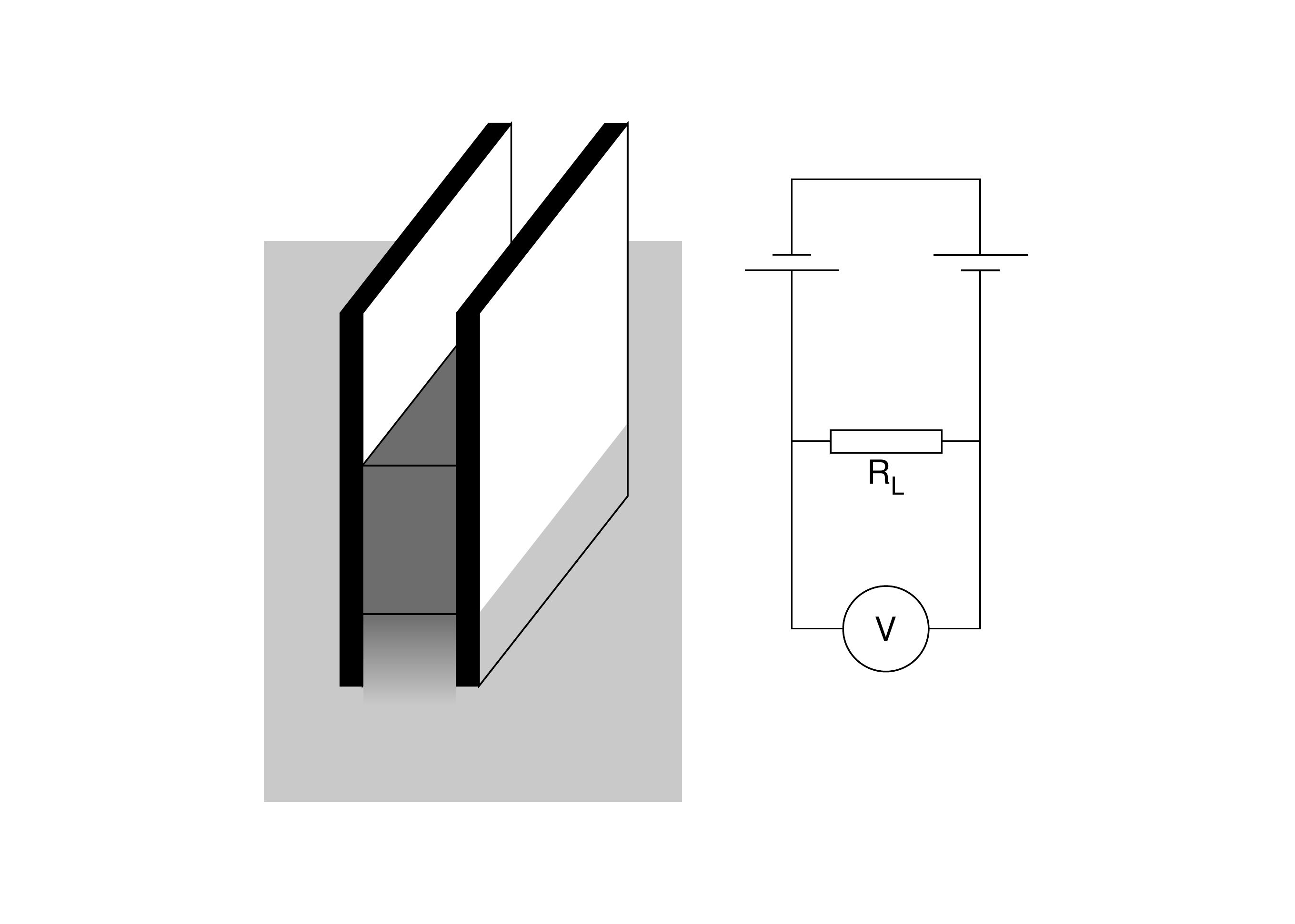}

\vspace{-2.2cm}

\caption{Left: View of the particle trap made of glass fibers, sandwiched between
two thermoelectric generators (TEGs) with efficiency 0.058~V/K, each and immersed
into KMnO$_4$ solution. Right: Circuit diagram with the TEGs represented as
battery symbols, for polarities see text, as well as voltmeter and load resistor.
\label{fig1}}
\end{figure}

Expectedly, the geometrical conditions inside the plug are crucial to motion
and density of Brownian particles therein, namely the two types of hydrated ions
in the KMnO$_4$ solution. As far as this can be manufactured, the fibers
are aligned vertically and at bottom and top of the plug they are
cut off. Clearly, a too dense plug handicaps the particles (thus retards their
diffusion) too strongly. As a rule of thumb, only around 5\% of the
entire plug consist of massive glass. Most conveniently, the plug is assembled
by an array of small parallel bundles, since otherwise capillary forces tend to
cluster the fibers into dense regions, while emptying the rest. Altogether, the
initial state of the experiment is the plug of glass fibers (framed by
the two TEGs) soaked by capillary forces with saturated KMnO$_4$ solution.

Further, Fig.~1 presents also the circuit diagram and, for the sake of
readability, we use battery symbols for the TEGs. The series circuit of the two
TEGs is performed such that both of them measure a temperature difference
between inside the plug and the surrounding liquid with same polarity
in their thermovoltages (not visible in the figure). Hence, the voltmeter displays
their sum $U$, and the sign is chosen to be positive if the plug is warmer than
around it. The load resistor $R_L$ equals 6.8~$\Omega$ which is twice the
internal resistance of a TEG, and so the device operates at maximum power
consumption $U^2/R_L$.

\section{Measurements and results}
The actual experiment consists of recurrent flushing of the plug from above, and
in the meantime diffusion (or Brownian motion) refills the plug to an amount that
is just adequate. All the time we measure the temperature difference between plug
and surrounding liquid in the container, that is essentially the thermal effects
within the plug, see Fig.~2. First we outline the technical procedure, and
afterwards we turn to the outcome. The very first flush at $t = 0$ in the
top row panels is deficient due to the saturated solution initially
present in the plug, which poorly matches the other system parameters.
Amazingly, the amount of flushing water of 8~cm$^3$ is a comparatively
uncritical setting. The one crucial point is to organize the totality
of experimental parameters in a way that the particle density gradient
between just flushed region (plug and favorably somewhat beyond) and
adjacent saturated solution is sufficiently steep. This turns out to be quite
an elaborate procedure of trial and error along gradients of improvement. A too
blurred zone is inadmissible, but we get by with a slightly inhomogeneous plug
that only in some areas by chance meets all the requirements. However, the
allowable volume in parameter space is tiny. In our view, to further enhance
reproducibility we should find appropriate "open" polymers whose chains replace
the glass fibers, but this then would turn into a really costly enterprise.

To the left in all our curves we see the lowermost part of a peak that is
by far the strongest thermal effect, namely warming due to flushing away
the evaporatively cooled surface water from the (rough) top of the plug. The
system parameters must be such that this peak is over rather quickly, since
the subsequent structure on the right-hand side in the upper panels is now
the effect we are after. As the Brownian particles enter the trap,
they lose a small percentage of their (at least translatory) kinetic energy
in favor of the host medium that warms up. The particles are now in
a quasi-equilibrium with the water molecules, and so equipartition is
ill-defined since there is no temperature that strictly applies to all
of the liquid. This situation persists as the particles diffuse further
into the trap, while new ones enter underneath. The spontaneous transition
from equilibrium to a slight deviation thereof entails an entropy
sink. Note that only the bottom region contributes to heating,
which renders conversion of voltages into temperatures somewhat dangerous.
Moreover, the high fluid level in the container (or deep immersion of the
TEGs, see above) raises the stability and thus improves the signal to noise ratio.

\begin{figure}

\vspace{-3.5cm}

\hspace{2.0cm}\includegraphics[scale=0.62, angle=0]{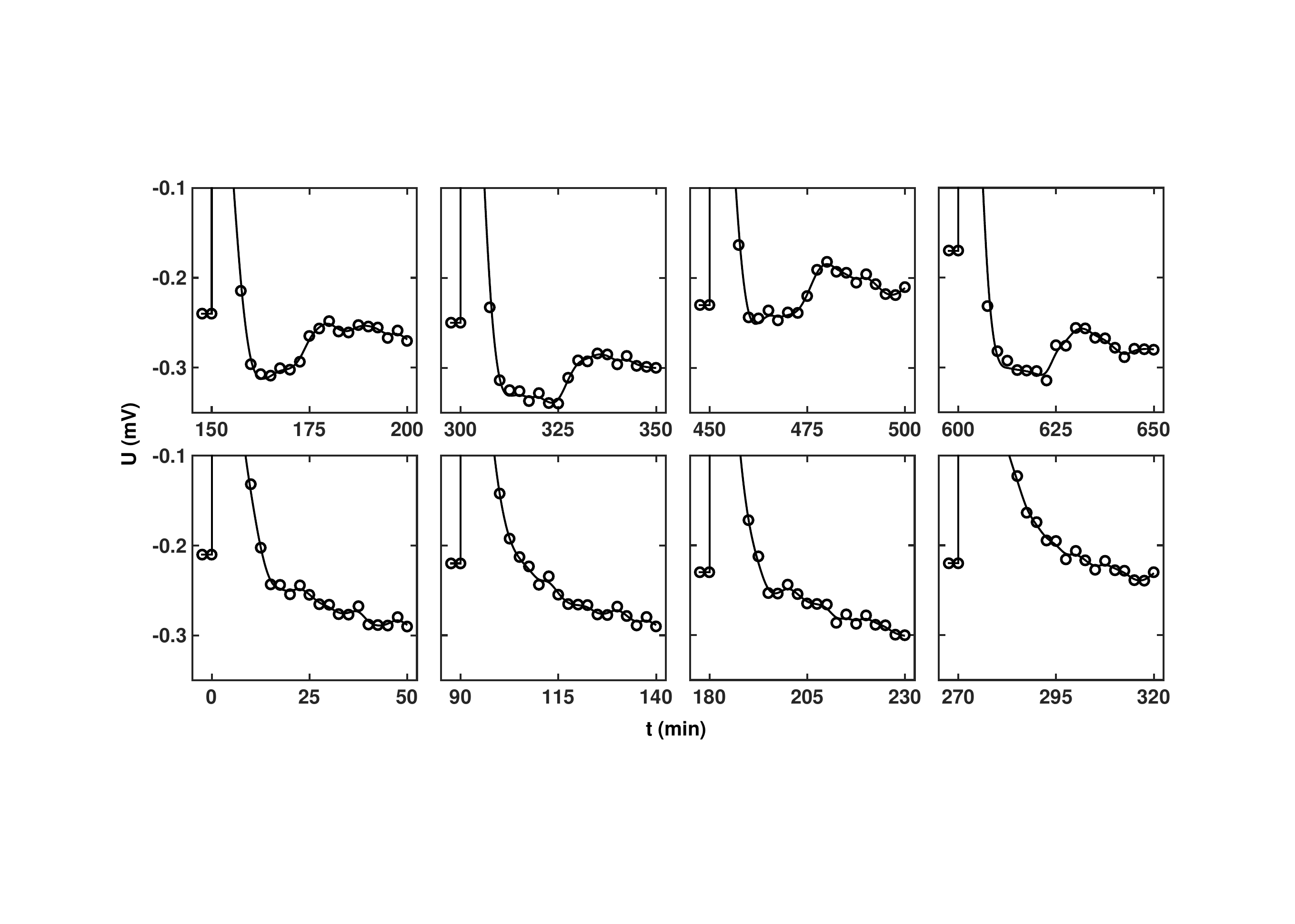}

\vspace{-3.0cm}

\caption{Top row: The rise at some 25~min after flushing issues from an energy
loss of the Brownian particles in favor of the host medium, caused by spontaneous
deviation from thermodynamic equilibrium of all the particles in the liquid.
Only the bottom region of the trap (particle inlet) brings about heating, and
so the plotted voltages by far underestimate the strength of this mechanism.
Bottom panels: In plain water, the suggested effect has gone. Apart from that,
the lasting descent within the plotted time windows is owing to different thermal
conditions such as evaporation rate in the absence of dissolved particles.
Regarding voltage measurement, the precision lies well within $\pm 0.01$~mV
throughout the figure.
\label{fig2}}
\end{figure}

\begin{figure}

\vspace{-4.0cm}

\hspace{2.0cm}\includegraphics[scale=0.62, angle=0]{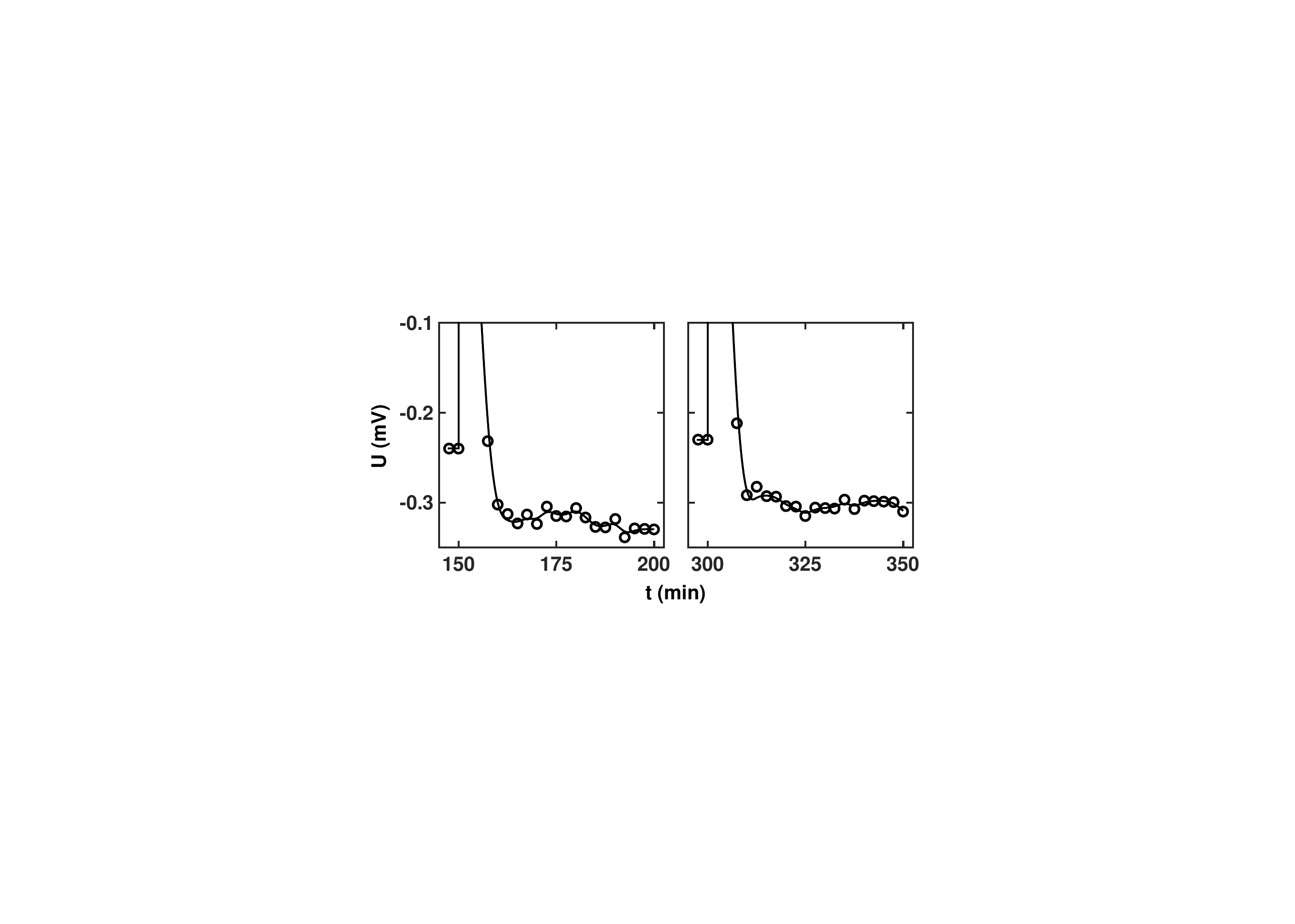}

\vspace{-4.8cm}

\caption{As in the upper panels of Fig.~2 but this time the plug of glass fibers
just slightly has missed the geometrical conditions for the desired diffusion
profile as a function of time. Hence, the particle flow into the trap was
insufficient, which was clearly visible to the unaided eye. The plots show little
effect, if at all.
\label{fig3}}
\end{figure}

For comparison, we may look at the four bottom panels in Fig.~2 where the effect
in question has disappeared. There we used the very same plug after replacing
the KMnO$_4$ solution by plain water. Just for readability we reset the
time scale to zero (actually zero means 14~hours in the upper scale). Here there
is no reason to omit the first run, and the intervals between consecutive plots
are shorter since we do not need to wait until diffusion has refilled the plug.
Further, in Fig.~3 we display the same situation as in the upper panels of
Fig.~2 (thus with KMnO$_4$ solution) using another plug that almost looked alike.
Nevertheless, it obviously has missed the requirements. We recognize similar plots
as in the top row of Fig.~2 but simply without the increase that is the outcome
of our endeavors. The desirable effect is absent, maybe up to an insignificant
hint thereof.

\section{Theory: the concept and models}
In order to substantiate our proposed mechanism of a marginal deviation from
thermodynamic equilibrium between Brownian particles and host medium, we
first recall some basics of Brownian motion and diffusion. Then we discuss
the all-important differences thereof. We state the Langevin equation for
a particle
\begin{equation}
m\ddot{{\bf x}}(t) = -\beta\dot{{\bf x}}(t)+{\bf F}_c(t)+{\bf F}_{ext}
\end{equation}
that, besides friction term and possible external force, comprises the random
driving force ${\bf F}_c(t)$ due to collision. Frequently, external
force means gravitation, but here this is negligible since the hydrated ions of
KMnO$_4$ do not sediment to the ground. The collision force is thought to be
a long-term function of time, i.e., an irregular sequence of many different
collisions.

This may be newly considered as a sum over many particles, as soon as we accept
that ergodicity applies in this case. Conveniently, isotropy removes the random
term from this sum, and as a final prerequisite we make use of equipartition
that introduces temperature. This way, after some steps of rearrangement
we arrive at our preferred version of Eq.~(1), namely an equation that
also covers the few just above stated clear-cut assumptions. This reads
\begin{equation}
\frac{d^2}{dt^2} \langle {\bf x}(t)^2 \rangle_x
+ \frac{\beta}{m} \frac{d}{dt} \langle {\bf x}(t)^2 \rangle_x
= \frac{2k_B T}{m},
\end{equation}
where the spatial mean value of squared distances the particles have departed
from their starting point is our new variable. The stationary solution of
Eq.~(2) is a fairly known result, namely the linearly growing mean value
\begin{equation}
\langle {\bf x}(t)^2 \rangle_x = 2Dt, \qquad D = \frac{k_B T}{\beta}.
\end{equation}
Here, $D$ is the diffusion constant as it enters the two major statements for
unconstrained diffusion
\begin{equation}
{\bf j}_n({\bf x},t) = -D\nabla n({\bf x},t), \qquad
\frac{\partial n({\bf x},t)}{\partial t} = D\Delta n({\bf x},t).
\end{equation}
The vector ${\bf j}_n({\bf x},t)$ denotes the particle current density, and
Eqs.~(4) describe Brownian motion as an overall diffusion phenomenon of many
particles with density $n({\bf x},t)$. We recognize that already unrestricted
diffusion constitutes a quasi-equilibrium of Brownian particles with the host
medium, since there exists a macroscopic time-evolution while the density
gradients tend to even out. Admittedly the process is slow, and so the
equipartition law stays almost unaffected.

However, we aim at a somewhat stronger deviation of the particles from equilibrium
with the background medium, although in terms of equilibrium energy this always
remains a minor percentage. Ref.~\cite{lancon} shows that the diffusion constant
$D$ strongly depends on limiting geometrical conditions, comparable to the
cavities in our plug of glass wool. The particles used by these authors
are somewhat larger but, since \cite{lancon} provides an experiment, the findings
there do not depend on the choice of theories we might consider to be
appropriate. In Ref.~\cite{moserarxiv} we performed a preliminary experiment, also
based on dissolved KMnO$_4$ and its diffusion within glass wool. There we
report on benefits of this choice, among them the exceptionally strong color
(of the MnO$_4$ ions) that permits visual inspection of particle density
and its propagation velocity inside the glass fiber environment. However,
the setup in \cite{moserarxiv} suffers from an unrecoverable shortcoming,
namely the single-step approach that cannot be upgraded to perpetual operation.
Thus, the quasi-equilibrium between Brownian particles and host medium may always
be taken as a transient regime prior to full equilibrium.

This problem is settled now. Our new device admits recurrent flushing that enables
us to repeatedly harvest the energy attended with the rise in the upper curves
of Fig.~2. Outside the trap, the equipartition energy of the Brownian particles
gets restored. Hence, there is an ongoing heat transfer from the container
into the trap. This transfer is not at all a consequence of thermal
equilibration that is much slower, and we may well bias the thermovoltages such
that the achieved increase starts at zero. We emphasize that this rise is the
issue where the second law gets violated, and so the various thermal interactions
with environment are immaterial. They affect just the temperature level (ordinates
in Fig.~2) where the warming happens, and we do not supply any thermal insulation.

The gained (or actually transferred) energy may be used now to organize flushing.
But this poses merely an engineering task, and by no means a matter of
principle. The flushing water may (optionally) be recycled by prior evaporation
out of the container. Again, the associated thermal interactions with the
environment just influence amount and temperature gradient of the heat backflow
from there. Note that, within the realm of the second law, such a delayed return
of energy borrowed from the surroundings is not admitted.

An engine designed as outlined above will indeed {\it run perpetually}. Its
efficiency $\eta$ fulfills $\eta_C < \eta \ll 1$ where the Carnot efficiency
$\eta_C$ is tiny since all occurring temperature differences (apart from
the initial peaks in Fig.~2) are minor. Based on the efficiency 0.058~V/K of the
used commercial TEGs and on geometry considerations, the relevant temperature
increase due to our proposed mechanism is in the order of 0.1~K. Probably
the limit of low efficiency has not so much been investigated, since generally
we aim at engines as efficient as possible. However, we think it is not a surprise
that close to this limit the crucial inequality $\eta > \eta_C$ more likely gets
a chance for a practical implementation.

Entropic phenomena caused by particle size are not new, depletion forces are
a long-known example, see Refs.~\cite{mao,karzar}. In order to model
the energy transfer from the Brownian particles to the smaller ones of the
background medium inside the porous vicinity, we suggest to rewrite the diffusion
constant $D$ in Eqs.~(3) as
\begin{equation}
\tilde{D} = \frac{k_B T}{\beta(d,r_p) + \tilde{\beta}(d,r_p,r_h)}
\end{equation}
where we purposely modify the coefficient $\beta$ of the dissipative term in
Eq.~(1). The distance $d$ in (5) is the typical spacing between the glass fibers,
and $r_p$ means particle radius (that optionally may be split up into
two different ones according to the unequal types of hydrated ions).
Then, $r_h$ denotes the typical radial extent of a host medium
particle. The first term $\beta(d,r_p)$ in the denominator of Eq.~(5)
is inspired by Ref.~\cite{lancon}, namely by the relative confinement
of Brownian particles that strongly enters the diffusion constant.
In addition, the term $\tilde{\beta}(d,r_p,r_h)$ accommodates also
$r_h$. This ansatz may well prove adequate, since the measured energy transfer
to the host medium is hard to imagine if we suppose all the particles to be
indistinguishable. We rule out a host or background medium that behaves just as
a continuum, and its smaller constituents are less inhibited in motion by the
cavity size. Moreover, an approach in the spirit of a generalized Fick-Jacobs
equation \cite{reguera,wang} would be rather inconvenient for our geometry of
aligned glass fibers (see above). Thus we stick to Eqs.~(4) and~(5), although
we are quite far from a suitable parametrization of (5). But, once this is
accomplished, we think the observed heating effect is primarily a function
of $\tilde{\beta}(d,r_p,r_h)$.

Thermal fluctuations, above all Brownian ones are sometimes said to be crucial
to generally explain the occurrence of irreversibility, not least because Eq.~(1)
is dissipative. We point to Jarzynski's equality
\begin{equation}
\langle e^{-W/k_B T} \rangle = e^{-\Delta F/k_B T}
\end{equation}
that relates the Helmholtz free energy difference $\Delta F$ to the work $W$
supplied to (or extracted from) a system, regardless wether or not we slowly
pursue a reversible path of quasi-static states starting at equilibrium situation
of reservoir temperature $T$ \cite{jarzynski}. Here, in $F = U - TS$ with $U$
being the internal energy, the dissipation-caused term $TS$ of entropy $S$
well may grow. The inequality (entropy remains or rises, or also
$\langle W \rangle \ge \Delta F$) is removed, and this supports strict validity
of the second law that this way might relate to other physical statemens.

However, we focus on a somewhat different issue, namely thermodynamics of
small systems, Refs.~\cite{sekimoto,wang2,brandner} meet our purposes.
The cavities in our particle trap are such limited systems that just weakly
couple to their environment, i.e., largely to adjacent hollows. Already
the constrained Brownian fluctuations inherently comprise all the properties
needed for the (marginal) second law violation in Fig.~2.

Likewise, we may look at Brownian fluctuations within the framework of linear
response theory. Suppose $x(t)$ is a (here scalar) unperturbed system's signal.
Instead, we state the progressing time-average of its response to a generalized
scalar external force $f(t)$ as
\begin{equation}
\langle x_{resp}(t) \rangle_t = \int^\infty_{-\infty} \Theta(t-\tau)
\chi(t-\tau) f(\tau) d\tau.
\end{equation}
Then, the Fourier transform $\tilde{\chi}(\omega)$ of the linear response function
or susceptibility $\chi(t)$ enters
\begin{equation}
\tilde{P}(\omega) = \frac{2k_B T}{\omega} \rm{Im} \tilde{\chi}(\omega)
\end{equation}
that is the fluctuation-dissipation theorem, see \cite{kubo,blundell}.
The Heaviside function $\Theta(t-\tau)$ in Eq.~(7) assures causality with respect
to the effect of $f(t)$, and $\tilde{P}(\omega)$ in (8) means the Fourier
transform of $x(t)^2$ or power spectral density. In a second law context,
causality is of utmost importance since we deal with the physics of a spontaneous
departure from strict equilibrium, very much in contrast to fictitious time
reversal in an irreversible process.

For a Brownian particle, theorem (8) states that the dissipation caused by
$f(t)$, thus essentially $\rm{Im} \tilde{\chi}(\omega)$, has the same physical
origin as the Brownian fluctuations (characterized by the $\tilde{P}(\omega)$
distribution) themselves. The cavity size or rather walls introduce now such an
$f(t)$ that selectively acts on particle size. Again, this substantiates the
marginal deviation from the subtle properties of true equilibrium dynamics.

We complete our choice of theoretical fundamentals and models with
a regard to quantum Brownian motion \cite{haenggi,bai}. But, instead of
replacing classical quantities by their quantum mechanical analogues
(such as momentum operators), we scrutinize those quantum trajectories
(for electrons) that exhibit physically correct velocities. These
velocities, or at least their magnitudes, are indispensable if we
aim at plain assignment to "unambiguous" thermodynamics. The key point
is consistency with wave mechanics, see \cite{ashcroft} for derivation
and numerical values of velocities, momenta, and energies in Sommerfeld's
theory of metals. In Ref.~\cite{moserannals} we present actual trajectories
(for one or more electrons in atoms) that, by virtue of position
vectors ${\bf x}(t)$, even provide angular momenta. Up to some poorly
surmountable numerical hurdles, they agree with wave mechanics. On this
footing, we think of an electron gas theory that more than so far will overcome
the gap between classical and quantum mechanical many-body theories. Again,
this would greatly enlarge the reach of the second law and of its exceptions.

\section{Conclusions}
In sum, we investigate the dependence of dynamics in a liquid on particle size
in a geometrically constrained environment. Brownian motion of larger particles
immersed into a background medium of smaller ones is an obvious candidate for this
intention. This way, the geometrical structure of a confinement matters, while
for point-like particles any cavity is of infinite relative extent. Then, we
organize the overall conditions of a diffusion experiment such that unequally
sized particles no longer carry the same equipartition energy. By virtue of
differently strong inhibition, they spontaneously adopt marginally deviant
kinetic energies, which establishes a quasi-equilibrium. It is of utmost
importance to recognize that even in the long run this situation does not
tend to the "usual" thermodynamic equilibrium. Hence, there is no temperature
that strictly holds for all of the particles. In terms of equilibrium energy,
the deviations thereof are minor but, nevertheless, they constitute a second
law exception.

We outline and, as far as possible with our present setup, quantitatively
evaluate what this means in terms of efficiency. Undoubtedly it is much easier
to design an arrangement that exceeds Carnot efficiency if we approach the limit
of zero efficiency. In this contribution we shall not scrutinize what measures
might improve the performance, and where this has fundamental limits. Admittedly,
automatized recurrent flushing of the particle trap is still missing, since on an
engineering level this cannot be carried out readily. But we strongly substantiate
that this will never destroy our envisaged entropy (and energy) balance. Our work
primarily covers entropic implications of size effects (cavities, particles),
and there we present results and viewpoints that, to our knowledge, are new.
In particular, we make use of these size effects in a pertinent way that unveils
a second law violation. That is what we have introduced by means of a particular
experiment and, in our view, explained in a retraceable manner. Thus we present
here a proof of concept, but the general situation is far from being settled.

\vspace{8mm}
\noindent
{\bf Acknowledgment}

The author thanks R. Maier, L. Pauli, P. Robmann, and U. Straumann for
constructive discussions.

\end{document}